\documentclass[twocolumn,showpacs,preprintnumbers,amsmath,amssymb]{revtex4}

\usepackage{times}
\usepackage{natbib}
\usepackage{amssymb,amsbsy,amsmath,amsfonts}
\usepackage{graphicx}
\usepackage{epsf,epsfig,float,latexsym,amsthm,fancyhdr,rotating}
\usepackage{graphics,psfrag,longtable}
\begin{document}
\title {The chiral representation of the $\pi N$ scattering amplitude and the pion-nucleon sigma term}
\author{J.M. Alarc\'on$^{1}$}
\author{J. Martin Camalich$^2$}
\author{J.A. Oller$^{1}$}
\affiliation{$^1$Departamento de F\'isica. Universidad de Murcia. E-30071, Murcia, Spain\\
$^2$Department of Physics and Astronomy, University of Sussex, BN1 9QH, Brighton, UK}
\begin{abstract}
We present a novel analysis of the $\pi N$ scattering amplitude in Lorentz covariant baryon chiral perturbation theory renormalized in the extended-on-mass-shell scheme. This amplitude, valid up to $\mathcal{O}(p^3)$ in the chiral expansion, systematically includes the effects of the $\Delta(1232)$ in the $\delta$-counting, has the right analytic properties and is renormalization-scale independent. This approach overcomes the limitations that previous chiral analyses of the $\pi N$ scattering amplitude had, providing an accurate description of the partial wave phase shifts of the Karlsruhe-Helsinki and George-Washington groups up to energies just below the resonance region. We also study the solution of the Matsinos group which focuses on the parameterization of the data at low energies. Once the values of the low-energy constants are determined by adjusting the center-of-mass energy dependence of the amplitude to the scattering data, we obtain predictions on different observables. In particular, we extract an accurate value for the pion-nucleon sigma term, $ \sigma_{\pi N}$. This allows us to avoid the usual method of extrapolation to the unphysical region of the amplitude. Our study indicates that the inclusion of modern meson-factory and pionic-atom data favors relatively large values of the sigma term. We report the value $\sigma_{\pi N}=59(7)$~MeV and comment on implications that this result may have.     

\end{abstract}
\pacs{13.75.Gx, 11.30.Rd, 12.39.Fe, 13.85.Dz}
\date{\today}
\maketitle

The sigma terms, $\sigma_{\pi N}$ and  $\sigma_s$, are observables of fundamental importance that embody the internal scalar structure of the nucleon, becoming an essential piece to understand the origin of the mass of the ordinary matter. The pion-nucleon sigma term, $\sigma_{\pi N}$, is a key ingredient in investigations of the QCD phase diagram and in the study of nuclear systems~\cite{NM:EOS,NM:ChRest}. On the other hand, $\sigma_{\pi N}$ and  $\sigma_s$ appear as hadronic matrix elements in the neutralino-nucleon elastic scattering cross section. Unfortunately, our current knowledge of the sigma terms is far from satisfactory. With the advent of experimental results on dark-matter searches, different authors have pled for a more accurate experimental determination of these quantities~\cite{Bottinoetal,Ellis:2008hf,Giedt:2009mr}.

The $\sigma_{\pi N}$ is defined as the nucleon matrix element of the light-quark scalar current,
\begin{equation}
\sigma_{\pi N}=\frac{1}{2 M_N}\langle N|\hat{m}\left(\bar{u}u+\bar{d}d\right)|N\rangle, \label{Eq:Sigmaterm-defintion} 
\end{equation}
where $\hat{m}=(m_u+m_d)/2$, with $m_u$, $m_d$ the light-quark masses. The sigma term can be obtained from the $\pi N$ scattering data by extrapolating the scattering amplitude to the Cheng-Dashen point~\cite{Cheng:1970mx,Hohler:1982ja}, which lies in the unphysical region of the Mandelstam plane. The usual method to perform this extrapolation is by means of an energy-dependent parameterization of the data in partial waves (PW) supplemented by dispersion relations that impose strong analyticity and unitarity constraints onto the scattering amplitude at low energies. The current uncertainty in $\sigma_{\pi N}$ originates from discrepancies between the classical PW analysis of the Karlsruhe-Helsinki (KH)~\cite{ka85} group and the more modern one performed by the George-Washington~\cite{wi08} (GW) group. More precisely, the KH amplitudes were used by Gasser \textit{et al.} to obtain the canonical result $\sigma_{\pi N}\simeq 45$~MeV~\cite{Gasser:1990ce}, whereas the analysis of the GW group, which includes modern meson factory data, leads to a larger value $\sigma_{\pi N}=64(7)$~MeV~\cite{Pavan:2001wz}. 

The main difficulty of the traditional method to obtain $\sigma_{\pi N}$ is assessing the errors that propagate in the extrapolation to the Cheng-Dashen point from the systematic uncertainties associated to a particular parametrization of the data. These problems, together with the persisting discrepancy in the values reported by the different PW analyses, have led to new strategies for the determination of the sigma terms. Particularly noteworthy is the intense campaign developed by the LQCD community to calculate these matrix elements using new powerful algorithms and computational resources~\cite{Young:2009ps}. 

In this paper, we focus on the extraction of the $\sigma_{\pi N}$ from $\pi N$ scattering data and using chiral perturbation theory ($\chi$PT), which is the effective field theory of QCD at low energies~\cite{ChPT,Bernard:1995dp,Scherer:2002tk,BernardII}. This is a suitable framework to shed light on the experimental discrepancies since it allows for an investigation of the chiral Ward identity that relates the isoscalar $\pi N$ scattering amplitude and $\sigma_{\pi N}$, giving a handle on the errors committed at each order of the power counting. In fact, one recovers the low-energy theorem at the Cheng-Dashen point that is exploited by the dispersive methods mentioned above. 

In $\chi$PT, one can alternatively use a more elegant manifestation of the same Ward identity between the two observables~\cite{Bernard:1995dp}. At ${\cal O}(p^3)$, $\sigma_{\pi N}$ only depends on one apriori unknown low-energy constant (LEC), $c_1$ (cf. Eq.~(2) below). Due to the non-linear realization of chiral symmetry underpinning $\chi$PT, this LEC also contributes to nucleon processes with an even number of external pion legs and, in particular, to the isoscalar part of the $\pi N$ scattering amplitude. Therefore, determining the value of this constant with a fit to the scattering data allows to predict $\sigma_{\pi N}$, avoiding any analytical extrapolation of the amplitude onto the unphysical region (as the one to the Cheng-Dashen point used in the dispersive analyses).

The low-energy structure of the $\pi N$ scattering amplitude has been studied within different approaches tackling the subtleties in the power counting that appear in the baryon sector of $\chi$PT (B$\chi$PT) (for reviews see Refs.~\cite{Bernard:1995dp,Scherer:2002tk,BernardII}). After the seminal paper of Gasser \textit{et al.}~\cite{Gasser:1987rb}, it was first studied in heavy-baryon (HB)  $\chi$PT~\cite{Jenkins:1990jv} by Fettes {\it et al.} up to $\mathcal{O}(p^3)$~\cite{Fettes:1998ud} and $\mathcal{O}(p^4)$~\cite{Fettes:2000xg} in the chiral counting. In these works, a precise description of the PWs was obtained at low energies, although the values of the LECs contain important contributions from the $\Delta(1232)$ resonance and the results for the $\sigma_{\pi N}$ were not accurate, being typically too large. The inclusion of the $\Delta$ as an explicit degree of freedom in the so-called small-scale-expansion (SSE)~\cite{Hemmert:1997ye} (that counts $\epsilon=M_\Delta-M_N\sim p$) up to $\mathcal{O}(\epsilon^3)$~\cite{FettesD}, offers a noticeably increase in the range of energies described compared with HB$\chi$PT at ${\cal O}(p^3)$~\cite{Fettes:1998ud}. Nonetheless, there is a strong dependence on the fitted values of the LECs with the PW analysis used as input that prevents a direct extraction of $\sigma_{\pi N}$ by fitting scattering data~\cite{FettesD}. After these difficulties, the conclusion was that the chiral convergence was not fast enough in the physical region so to extract useful information on $\sigma_{\pi N}$ from the PW phase shifts~\cite{Buettiker:1999ap}. 

It has been shown that the non-relativistic expansion implemented in the HB approach does not converge in part of the low energy region ~\cite{Bernard:1995dp,Becher:1999he,Fuchs:2003kq}. This led to the studies in the manifestly Lorentz covariant infrared (IR) B$\chi$PT~\cite{Becher:1999he,Becher:2001hv,IREllis,IROurs}. In this case, the amplitude up to $\mathcal{O}(p^4)$ without the $\Delta$ as explicit degree of freedom, shows an accurate and rapidly convergent description in the subthreshold region but fails to connect it to the physical one~\cite{Becher:2001hv}, confirming the conclusions about $\sigma_{\pi N}$ drawn from the previous works in HB. 

\begin{table}
\centering
\caption{Physical observables obtained from the $\mathcal{O}(p^3)$ $\pi N$ scattering amplitude in the EOMS renormalization scheme fitted to different PW analyses up to $W_{max}=1.2$~GeV ($W_{max}=1.16$~GeV for EM). The scattering lengths are in units of 10$^{-2}$ $m_\pi^{-1}$.\label{Table:Observables}}
\begin{ruledtabular}
\begin{tabular}{ccccccc}
&$\chi^2_{\rm d.o.f.}$&$h_A$&$g_{\pi N}$&$\Delta_{GT}$ [\%] &$a_{0+}^+$ &$a_{0+}^-$ \\
\hline
KH~\cite{ka85}&0.75&3.02(4)&13.51(10)&4.9(8)&$-1.2(8)$&8.7(2)\\
GW~\cite{wi08}&0.23&2.87(4)&13.15(10)&2.1(8)&$-0.4(7)$&8.2(2)\\
EM~\cite{Matsinos:2006sw}&0.11&2.99(2)&13.12(5)&1.9(4)&0.2(3)&7.7(1)
\end{tabular}
\end{ruledtabular}
\end{table}

In this paper, we present a $\chi$PT analysis of the $\pi N$-scattering amplitude and of the pion-nucleon sigma term up to $\mathcal{O}(p^3)$ accuracy that includes two main improvements over previous work. In the first place, we use Lorentz covariant B$\chi$PT with a consistent power counting obtained via the extended-on-mass-shell (EOMS) renormalization scheme~\cite{EOMS}. This prescription, instead of IR, is used because the latter introduces unphysical cuts that may influence the low-energy region~\cite{Holstein,Geng:2008mf,BernardII}. As it has been recently shown in Ref.~\cite{piNLarge}, the huge Goldberger-Treiman (GT) discrepancy, of $\sim20\%$, found in the IR scheme~\cite{IREllis,IROurs} can be traced back to the analyticity issues of this method rather than to a breaking of the chiral convergence in the $\pi N$ system. In addition, we obtain amplitudes independent of the renormalization scale, which is not the case for those given by IR~\cite{Becher:1999he,Becher:2001hv,IREllis,IROurs}. Secondly, we explicitly include the $\Delta$ taking into account that, below the resonance region, the diagrams with the $\Delta$ are suppressed in comparison with those with the nucleon. This can be implemented in the so-called $\delta$-counting by assigning an extra fractional suppression of $\mathcal{O}(p^{1/2})$ to the $\Delta$-propagators in the Feynman diagrams~\cite{Pascalutsa,PascalutsaII}. For the $N\Delta$ chiral Lagrangians, we use the $consistent$ formulation of Pascalutsa~\cite{Pascalutsa:2000kd,PascalutsaII,Krebs:2008zb} which filters the unphysical components of the relativistic spin-3/2 spinors and eliminate the dependence on off-shell parameters that the conventional vertices have. The technical details of this calculation and the complete results derived thereafter are presented in detail elsewhere~\cite{piNLarge}. In the following, we outline the analysis and show its main results on the $\pi N$ phase shifts and, more specifically, on the pion-nucleon sigma term.

\begin{figure}[t] 
\includegraphics[scale=1]{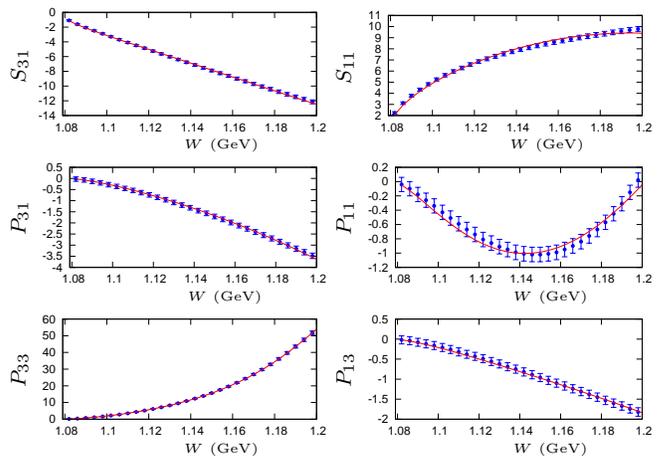} 
\caption{(Color on-line) Phase shifts given by the Lorentz covariant $\mathcal{O}(p^3)$ $\pi N$ scattering amplitude in the EOMS scheme fitted to the GW solution (circles)~\cite{wi08} up to $W_{max}=1.2$~GeV. \label{fig_graph}}
\end{figure}

The calculation proceeds as in Ref.~\cite{IROurs} but with the loops of that reference treated in the EOMS scheme. This is achieved by canceling the UV divergences obtained in dimensional regularization such that the power-counting breaking pieces of the loops are absorbed into the $\mathcal{O}(p)$ LECs, $g_A$ (axial coupling of the nucleon) and $M_N$, and into the 4 $\mathcal{O}(p^2)$ LECs, $c_1$, $c_2$, $c_3$, $c_4$. The 5 combinations of $\mathcal{O}(p^3)$ LECs, $d_1+d_2$, $d_3$, $d_5$, $d_{14}-d_{15}$ and $d_{18}$ are renormalized in the $\overline{MS}$ scheme.  Besides that, we also include the Born-term with an intermediate $\Delta(1232)$ resonance and leading $\mathcal{O}(p)$ vertices given by the $N\Delta$ axial coupling $h_A$. The Born-terms with  $\mathcal{O}(p^2)$ $N\Delta$ couplings~\cite{FettesD,Geng:2008bm,Long:2010kt} have also been considered but they give a negligible contribution and have been omitted in the present study, whereas the corresponding loops with $\Delta$ propagators are of higher-order. 

We fix the values of the LECs fitting the center-of-mass (CM) energy dependence of the 2 $S$- and 4 $P$-wave phase shifts obtained from the chiral amplitude to the latest solutions of the KH~\cite{ka85} and GW~\cite{wi08} groups. In addition, we include the analysis of the Matsinos' group (EM) ~\cite{Matsinos:2006sw} which focuses on the PW parameterization of the data at very low energies without imposing dispersive constraints from the high-energy region. We follow the logic of Ref.~\cite{IROurs} to assign errors to the first two analyses (they do not provide errors) while for the latter we include the errors provided there. The fits are done from the lowest CM energies above threshold, $W_{th}\simeq 1.08$~GeV, up to $W_{max}=1.2$~GeV which is below the $\Delta(1232)$ region (the EM analysis only reaches $W_{max}\simeq1.16$~GeV). The parameters $g_A$, $M_N$, $M_\Delta$, $m_\pi$ and $f_\pi$ are fixed to their experimental values~\cite{IROurs}. Although the $N\Delta$ axial coupling can be determined using the $\Delta(1232)$ width, we also fit this constant to the PW phase shifts. The suitable value to compare with is the one obtained from the Breit-Wigner width $\Gamma_\Delta=118(2)$~MeV~\cite{Nakamura:2010zzi}, namely $h_A=2.90(2)$~\cite{PascalutsaII}.

\begin{table}[h]
\centering
\caption{Values of the $\mathcal{O}$($p^2$) LECs in units of GeV$^{-1}$ and of $\sigma_{\pi N}$ in MeV obtained from the different $\pi N$ PW analyses.
 \label{Table:ResSigma}}
\begin{ruledtabular}
\begin{tabular}{cccccc}
&$c_1$ &$c_2$ &$c_3$ &$c_4$&$\sigma_{\pi N}$\\
\hline
KH&$-$0.80(6)&1.12(13)&$-$2.96(15)&2.00(7)&43(5)\\
GW&$-$1.00(4)&1.01(4)&$-$3.04(2)&2.02(1)&59(4)\\
EM&$-$1.00(1)&0.58(3)&$-$2.51(4)&1.77(2)&59(2)
\end{tabular}
\end{ruledtabular}
\end{table}

The results on a selection of physical observables are shown in Table~\ref{Table:Observables}. The errors quoted there are only of statistical origin and additional theoretical uncertainties are to be added. In Fig.~\ref{fig_graph} we plot the phase shifts of the $S$- and $P$-waves given by the $\pi N$ scattering amplitude in the EOMS scheme and at $\mathcal{O}(p^3)$ in the $\delta$-counting, fitted to the GW solution (circles)~\cite{wi08} up to $W_{max}=1.2$~GeV. Similar plots can be obtained for the KH and EM solutions. The figure shows that the description of the lowest PWs is very accurate up to energies below the $\Delta$-resonance region, covering a range of energies larger than in previous perturbative analyses~\cite{Fettes:2000xg,FettesD,IREllis,IROurs}. The quality of the description is reflected by the small $\chi^2_{\rm d.o.f.}$ listed in the second column of Table ~\ref{Table:Observables}, which furthermore shows that the description of GW and EM PW analyses is better than the KH one. 

As one can infer from the third column of Table~\ref{Table:Observables}, only the GW solution gives a result on $h_A$ that is perfectly compatible with the determination from the $\Delta$ width. In the fourth column, we show the values obtained for the $\pi N$ coupling that, compared with the axial coupling $g_A$, gives the GT discrepancy $\Delta_{GT}$ in the fifth column. These can be compared with the numbers independently extracted from $NN$-scattering ($g_{\pi N}\simeq13.0$)~\cite{deSwart:1997ep} and pionic atom ($g_{\pi N}=13.12(9)$)~\cite{Baru:2010xn} data. For the discussion of the large violation obtained in IR B$\chi$PT~\cite{IREllis,IROurs} as compared with the one obtained here, see Ref.~\cite{piNLarge}. Results for the isoscalar ($a_{0+}^+$) and isovector ($a_{0+}^-$) scattering lengths are shown in the last two columns of Table~\ref{Table:Observables}. Minding that changes of 5-10\% can be easily expected from higher-order and isospin corrections~\cite{Hoferichter:2009ez}, we can compare with the values independently extracted from pionic-atom data, $a_{0+}^-=0.0861(9)$ and $a_{0+}^+=0.0076(31)$ $m_\pi^{-1}$~\cite{Baru:2010xn}. The impact that the pionic-atom result for $a_{0+}^+$ has on the value of $\sigma_{\pi N}$~\cite{Gasser:1990ce,Olsson:1999jt} is addressed below.

As mentioned above, the isoscalar $\pi N$ scattering amplitude is related with $\sigma_{\pi N}$ through the LEC $c_1$. In the first columns of Table~\ref{Table:ResSigma}, we show the fitted values for the $\mathcal{O}(p^2)$ LECs. The values and errors quoted there correspond to the mean and standard deviation obtained after considering fits to the KH and GW PW phase shifts for various $W_{max}$, from $1.14$~GeV to $1.2$~GeV ($W_{max}\sim1.16$~GeV for the EM analysis) in intervals of $0.01$~GeV. The purpose of this strategy is to take into account the dispersion of the values of these LECs (and of $\sigma_{\pi N}$) against the data set included in the fits. As we can see, our results remain stable to the increase of the maximum energy and to the particular analysis used as experimental input. This is in remarkable contrast with the strong sensitivity of the values of the LECs obtained in the ${\cal O}(\epsilon^3)$ study done in HB$\chi$PT-SSE~\cite{FettesD}. 

On the other hand, the results in Table~\ref{Table:ResSigma} are quite different to the ones obtained without the explicit inclusion of the $\Delta$. In this case, we can  describe the GW phase shifts up to $W_{max}\simeq1.14$~GeV ($\chi^2_{\rm d.o.f.}=0.62$) and we obtain $c_1=-1.54(5)$, $c_2=3.92(6)$, $c_3=-6.87(6)$ and $c_4=3.79(3)$ (all in units of GeV$^{-1}$). Comparing with the values in Table~\ref{Table:ResSigma}, we see that the contribution of the $\Delta$ to the $\mathcal{O}(p^2)$ LECs $c_{2-4}$ is compatible with the one estimated by resonance saturation hypothesis~\cite{Bernard:1996gq}, $c_2^\Delta=1.9\ldots3.8$, $c_3^\Delta=-3.8\ldots-3$ and $c_4^\Delta=1.4\ldots2.0$ (in GeV$^{-1}$). For the $c_1$ counter-term the $\Delta$ contribution is negligible~\cite{Bernard:1996gq,Becher:1999he}. We interpret the difference, of around $0.5$~GeV$^{-1}$, between our result in the second row of Table~\ref{Table:ResSigma} including the $\Delta$ and that without this resonance, as a clear indication that the LECs are stabilized once the tree-level $\Delta$ exchange contributions are taken into account~\cite{Krebs:2007rh,Long:2010kt}. 

We calculate $\sigma_{\pi N}$ at ${\cal O}(p^3)$ employing covariant B$\chi$PT in the EOMS renormalization scheme. The pion-nucleon sigma term can be obtained either from the scalar form factor of the nucleon, Eq.~(\ref{Eq:Sigmaterm-defintion}), or from the quark mass dependence of its mass and the Hellmann-Feynman theorem. The resulting expression is
\begin{widetext}
\begin{eqnarray}
\sigma_{\pi N}=-4c_1 m_\pi^2-\frac{3g_A^2 m_\pi^3}{16\pi^2f_\pi^2 M_N}\left(\frac{3M_N^2-m_\pi^2}{\sqrt{4M_N^2-m_\pi^2}}\arccos{\frac{m_\pi}{2M_N}}
+m_\pi\log{\frac{m_\pi}{M_N}}\right),\label{Eq:SpiN}
\end{eqnarray}   
\end{widetext}   
and leads, in the non-relativistic limit, to the HB result up to $\mathcal{O}(p^3)$~\cite{Bernard:1993nj}. Besides the error propagated from $c_1$, this expression carries a theoretical uncertainty coming from higher-order contributions. We estimate this by computing the next subleading correction, at $\mathcal{O}(p^{7/2})$ in the $\delta$-counting, which is given by a loop diagram with an insertion of a $\Delta$ propagator~\cite{Pascalutsa:2005nd}. This amounts to a contribution of $-6$~MeV (that we take as an irreducible uncertainty in our determination) to be compared with the one at $\mathcal{O}(p^{3})$ of $-19$~MeV. Furthermore, we have calculated the $\mathcal{O}(p^{4})$ corrections given by the loop diagrams with an insertion of the $\mathcal{O}(p^2)$ LECs~\cite{Becher:1999he}. With the values of the LECs in Table~\ref{Table:ResSigma} we obtain that they span from $-2$ to $-4$ MeV. These results suggest a clear convergence pattern for the chiral expansion of $\sigma_{\pi N}$ as well as they confirm the hierarchy at low energies between the nucleon and $\Delta$ contributions that is implemented in the $\delta$-counting. 

With Eq.~(\ref{Eq:SpiN}) and the values for $c_1$ obtained in the interval $1.14$~GeV $\leq W_{max}\leq1.2$ GeV, we determine the means and standard deviations of $\sigma_{\pi N}$ listed in the last column of Table~\ref{Table:ResSigma}. The values of $\sigma_{\pi N}$ extracted from the different analysis are not completely consistent among each other. The KH number reproduces the canonical result $\sigma_{\pi N}\simeq45~$MeV~\cite{Gasser:1990ce}, whereas  those determined from the GW and EM solutions agree with the dispersive result of the GW group $\sigma_{\pi N}=64(7)$~MeV~\cite{Pavan:2001wz}. Furthermore, the result from the EM analysis also agrees with $\sigma_{\pi N}\simeq56(9)$~MeV, obtained by Olsson~\cite{Olsson:1999jt} using a dispersive sum-rule and the threshold parameters provided by an early version of the EM solution. 

Although our results for each of the PW solutions are consistent with those obtained extrapolating the data to the Cheng-Dashen point, the B$\chi$PT approach applied here relies solely on the information in the region where the data actually exist. We then give an estimation on the uncertainty committed in the relation between $\sigma_{\pi N}$ and the $\pi N$ scattering amplitude that is based on effective field theory grounds. On top of that, the dispersion of the results with respect $W_{max}$ and among the different analyses allows to disentangle the systematics coming either from the data basis employed or the particular parameterization of the data. In this sense, the consistency between the results derived from the GW and EM solutions is very remarkable since these are quite different PW parameterizations having both in common the inclusion of the wealth of low-energy data collected along the last 20 years in meson and pion  factories~\cite{wi08,Matsinos:2006sw} with many points not included in KH~\cite{ka85}. Therefore, our results suggest that the modern meson scattering data lead to a value for $\sigma_{\pi N}$ larger than the one obtained from the older KH analysis~\cite{Gasser:1990ce}. Nevertheless, a re-analysis of the modern data set with the KH method would be extremely valuable in order to reach a definite conclusion in this regard (see e.g. Ref.~\cite{Osmanovic:2011xn}).

Another important and independent source of information comes from the pionic-atom data on $a^+_{0+}$. It has been noted before in dispersive studies~\cite{Gasser:1990ce,Olsson:1999jt,Pavan:2001wz} that the sign of this observable is strongly correlated with the value of $\sigma_{\pi N}$. While the KH result is compatible with the old negative results, it is not anymore with the recent positive values extracted from modern pionic-atom data and using improved phenomenological approaches~\cite{Baru:2010xn}. These are, on the other hand, compatible with the scattering data determinations obtained from the GW and EM solutions. The effect that a non-negative result on $a^+_{0+}$ has on $\sigma_{\pi N}$ was quantitatively studied by the GW group concluding that a value of $a^+_{0+}\gtrsim 0$ produces a raise on the sigma term of, at least, 7~MeV~\cite{Pavan:2001wz}. 

Finally, we want to emphasize that only our results based in the GW analysis are perfectly compatible with all the phenomenology that can be extracted from independent experimental sources.  We remind here that the KH analysis gives rise to a value for $h_A$ that is not compatible with the value obtained from the $\Delta(1232)$ width (in agreement with the KH overestimation of this observable) and to a value for $g_{\pi N}$ that leads to a sizable violation of the GT relation, which is nowadays theoretically implausible.  As for our study of the EM PW analysis, we found a value for the isovector scattering length that is too small as compared with the accurate values obtained from pion-atoms data~\cite{Matsinos:2006sw,Baru:2010xn}.

With these considerations, one obtains the following value for $\sigma_{\pi N}$, as it is extracted from the analysis of $\pi N$ modern scattering data~\cite{wi08,Matsinos:2006sw} and using Lorentz covariant B$\chi$PT in the EOMS scheme up to $\mathcal{O}(p^3)$ in the $\delta$-counting,
\begin{equation}
\sigma_{\pi N}=59(7) {\rm MeV}.\label{Eq:SpiNValue}
\end{equation}
The error includes the higher-order uncertainty estimated above added in quadrature with the one given by the dispersion of the values in the average of the GW and EM results. If one were to include the KH result in this estimation, the result would be slightly reduced by 2-3~MeV.

As a concluding observation we want to address the fact that this relatively large value of $\sigma_{\pi N}$ may appear to be in conflict with some established phenomenology. In particular, it may give a new twist to the old puzzle concerning the strangeness content of the nucleon~\cite{Pavan:2001wz}. This is based on the relation that is obtained in HB$\chi$PT up to $\mathcal{O}(p^4)$ accuracy among the $SU(3)_F$-breaking of the baryon-octet masses, $\sigma_{\pi N}$ and the observable $y$ quantifying the strangeness content of the nucleon~\cite{Gasser:1982ap,Borasoy:1996bx}. For the value of the sigma term obtained in the present work, this relation leads to a contribution of the strange quark to the nucleon mass of several hundreds of MeV. It is interesting to note that the usual method to derive this relation does not include explicitly the effects of the decuplet resonances, which have been shown to largely cancel those of the octet in the strangeness content of the nucleon~\cite{Jenkins:1991bs,Bernard:1993nj}. This is, indeed, consistent with recent B$\chi$PT determinations of the sigma terms using LQCD results on the baryon masses and explicitly including the decuplet contributions~\cite{Young:2009zb,MartinCamalich:2010fp}, showing that a relatively large value of $\sigma_{\pi N}\simeq 60$ MeV~\cite{MartinCamalich:2010fp} is not at odds with a negligible strangeness in the nucleon. Another caveat arises in chiral approaches to nuclear matter, in which a large value of $\sigma_{\pi N}$ would lead to a vanishing quark condensate at too low densities~\cite{NM:EOS,NM:ChRest}. It is important to note that a non-zero value of the in-medium temporal component of the pion axial coupling, $f_t$, is also a necessary condition for the spontaneous breaking of chiral symmetry ~\cite{OllerMedium}. Hence, an analysis of the density dependence of this quantity, together with the quark condensate, is necessary in order to properly discuss about chiral symmetry restoration in nuclear matter.

In summary, we have presented a novel analysis of the $\pi N$ scattering amplitude in Lorentz covariant B$\chi$PT within the EOMS scheme up to ${\cal O}(p^3)$ and including the effects of the $\Delta(1232)$ explicitly in the $\delta$-counting. This covariant approach ensures the right analytic properties of the tree-level and loop corrections to the amplitude, providing a model-independent framework to comprehensively and accurately study the phenomenology associated with the different PW parameterizations of the $\pi N$-scattering data. In particular, we found that we perfectly describe the PW phase shifts of the KH, GW and EM groups up to energies below the $\Delta$-resonance region, at the same time as we agree in the values of the scattering observables. It is worth stressing that, apart from the phase shifts, our results using the GW analysis are perfectly compatible on important observables with those obtained from independent phenomenological sources.

We show that our amplitudes are suitable to extract an accurate value of $\sigma_{\pi N}$ from scattering data and avoiding the extrapolation to the unphysical region using a method based on EOMS-B$\chi$PT. Namely, the pion-nucleon sigma term can be properly calculated at ${\cal O}(p^3)$ only when B$\chi$PT is formulated in a way that keeps the analytical properties of the amplitude, accounts for the important effects of the $\Delta$ resonance in the LECs and gives results independent of the renormalization scale. It follows that the extraction method to calculate $\sigma_{\pi N}$ is under good theoretical control. Consequently, we ratify the discrepancy between the KH and GW groups and give support to the latter, which is in agreement with the one that we obtain from the study of the latest EM solution. We conclude that recent analyses of the modern data lead to a relatively high value of $\sigma_{\pi N}$, cf. Eq.~(\ref{Eq:SpiNValue}).

The authors want to acknowledge L.~Alvarez-Ruso, E.~Epelbaum, L.~S.~Geng, U.~G.~Meissner, V.~Pascalutsa and W.~Weise for reading the manuscript and for their useful comments. This work is funded by the grants FPA2010-17806 and the Fundaci\'on S\'eneca 11871/PI/09. We also thank the financial support from  the BMBF grant 06BN411, the EU-Research Infrastructure Integrating Activity ``Study of Strongly Interacting Matter" (HadronPhysics2, grant n. 227431) under the Seventh Framework Program of EU and the Consolider-Ingenio 2010 Programme CPAN (CSD2007-00042). JMC acknowledges the MEC contract FIS2006-03438, the EU Integrated Infrastructure Initiative Hadron Physics Project contract RII3-CT-2004-506078 and the STFC [grant number ST/H004661/1] for support.

\end{document}